\documentclass{mem}

\usepackage{natbib}
\usepackage{txfonts}
\usepackage{balance}
\usepackage{graphicx}
\usepackage[a4paper]{hyperref}
\idline{00}{189}

\def\ks{$\rm{km}~\rm{s}^{-1}$}
\def\ksk{$\rm{km}~\rm{s}^{-1}~\rm{kpc}^{-1}$}

\begin{document}

\title{New method for determining the Milky Way bar pattern speed}

\subtitle{}

\author{I. \,Minchev\inst{1}, J. \,Nordhaus\inst{2} \and
  A. C. \,Quillen\inst{3} }

\offprints{Ivan Minchev; \email{minchev@astro.u-strasbg.fr}}

\institute{Observatoire Astronomique, Universit\'e de Strasbourg, 
  Strasbourg, France
\and
  Department of Astrophysical Sciences, Princeton University, 
  Princeton, NJ, USA
\and
  Department of Physics and Astronomy, University of Rochester,
  Rochester, NY, USA
}

\authorrunning{Minchev, Nordhaus, and Quillen}

\titlerunning{Milky Way bar pattern speed}

\abstract{
Previous work has related the Galactic bar to structure in the local
stellar velocity distribution. Here we show that the bar also
influences the spatial gradients of the velocity vector via the Oort
constants.  By numerical integration of test-particles we simulate
measurements of the Oort $C$ value in a gravitational potential
including the Galactic bar. We account for the observed trend that $C$
is increasingly negative for stars with higher velocity dispersion.
By comparing measurements of $C$ with our simulations we improve on
previous models of the bar, estimating that the bar pattern speed is
$\Omega_{\rm b}/\Omega_0=1.87\pm0.04$, where $\Omega_0$ is the local
circular frequency, and the bar angle lies within
$20^\circ\leqslant\phi_0\leqslant45^\circ$.  We find that the Galactic
bar affects measurements of the Oort constants $A$ and $B$ less than
$\sim2$ \ksk\ for the hot stars.}  

\maketitle{}

\section{Introduction}

The Galaxy is often modeled as an axisymmetric disk.  With the ever
increasing proper motion and radial velocity data, it is now apparent
that non-axisymmetric effects cannot be neglected (nonzero Oort
constant $C$, \citealt{olling03}, hereafter OD03; nonzero vertex
deviation, \citealt{famaey05}; asymmetries in the local velocity
distribution of stars, \citealt{dehnen98}).  It is still not clear,
however, what the exact nature of the perturbing agent(s) in the Solar
neighborhood (SN) is(are).  Possible candidates are spiral density
waves, a central bar, and a triaxial halo.

Due to our position in the Galactic disk, the properties of the Milky
Way bar are hard to observe directly.  Hence its parameters, such as
orientation and pattern speed, have only been inferred indirectly from
observations of the inner Galaxy (e.g., \citealt{blitz91,weinberg92}).

However, the bar has also been found to affect the local velocity
distribution of stars. HIPPARCOS data revealed more clearly a stream
of old disk stars with an asymmetric drift of about 45 \ks\ and a
radial velocity $u<0$, with $u$ and $v$ positive toward the Galactic
center and in the direction of Galactic rotation, respectively.  This
agglomeration of stars has been dubbed the `Hercules' stream or the
`$u$-anomaly'. The numerical work of \cite{dehnen99,dehnen00} and
\cite{fux01} has shown that this stream can be explained as the effect
of the Milky Way bar if the Sun is placed just outside the 2:1 Outer
Lindblad Resonance (OLR). Using high-resolution spectra of nearby F
and G dwarf stars, \cite{bensby07} have investigated the detailed
abundance and age structure of the Hercules stream. Since this stream
is composed of stars of different ages and metallicities pertinent to
both thin and thick disks, they concluded that a dynamical effect,
such as the influence of the bar, is a more likely explanation than a
dispersed cluster.

Assuming the Galactic bar affects the shape of the distribution
function of the old stellar population in the SN, an additional
constraint on the bar can be provided by considering the values
derived for the Oort constant $C$. In other words, in addition to
relating the dynamical influence of the bar to the local velocity
field, $C$ provides a link to the gradients of the velocities as well.
The study of OD03 not only measured a non-zero $C$, implying the
presence of non-circular motion in the SN, but also found that $C$ is
more negative for older and redder stars with a larger velocity
dispersion. This is surprising as a hotter stellar population is
expected to have averaged properties more nearly axisymmetric and
hence, a reduced value of $|C|$ (e.g., \citealt{mq07}; hereafter Paper
I).

It is the aim of this work to show it is indeed possible to explain
the observationally deduced trend for $C$ (OD03) by modeling the Milky
Way as an exponential disk perturbed by a central bar.  By performing
this exercise, we provide additional constraints on the bar's pattern
speed. Models for the structure in the bulge of our galaxy are
difficult to constrain because of the large numbers of degrees of
freedom in bar models. Thus future studies of structure in the
Galactic Center will benefit from tighter constraints on the
parameters describing the bar, such as its pattern speed and angle
with respect to the Sun.

\section{The Oort constants}

\begin{table*}
\caption{Simulation parameters used\label{table:par}}
\begin{center}
\begin{tabular}{lcc}
\hline
\noalign{\smallskip}
Parameter & Symbol & Value  \\
\noalign{\smallskip}
\hline
\noalign{\smallskip}
Solar neighborhood radius   & $r_0$  &   1                             \\
Circular velocity at $r_0$  & $v_0$  &   1                             \\
Radial velocity dispersion & $\sigma_{u}(r_0)$ & $0.05v_0$ or $0.18v_0$ \\
$\sigma_{\rm u}$ scale length & $r_{\sigma}$ & $0.9r_0$                  \\
Disk scale length &  $r_\rho$       &    $0.37r_0$                      \\
Bar strength      &  $\epsilon_{\rm b}$   &     $-0.012$                \\
Bar size          &   $r_{\rm b}$         &     $0.8r_{\rm cr}$          \\ 
\noalign{\smallskip}
\hline
\end{tabular}
\end{center}
\end{table*}

We can linearize the local velocity field (e.g., Paper I) about the
LSR and write the mean radial velocity $\overline{v}_d$ and
longitudinal proper motion $\overline{\mu}_l$ as functions of the
Galactic longitude $l$ as
\begin{eqnarray}
\label{eq:vel}
{\overline{v}_d\over \overline{d}} &=& K + A\sin(2l) + C\cos(2l)   \\
\overline{\mu}_l &=& B + A\cos(2l) - C\sin(2l)\nonumber
\end{eqnarray}
where $\overline{d}$ is the average heliocentric distance of stars, $A$ and $B$
are
the usual Oort constants, and $C$ and $K$ are given by
\begin{eqnarray}
\label{eq:oc}
2C  &\equiv &   -{\overline{u} \over r} + {\partial \overline{u} \over \partial
r}
- {1\over r} {\partial \overline{v}_{\phi} \over \partial \phi}         \\
  2K  &\equiv &   +{\overline{u} \over r} + {\partial \overline{u} \over
\partial r}
+ {1\over r} {\partial \overline{v}_{\phi} \over \partial \phi}.
\end{eqnarray}
Here $r$ and $\phi$ are the usual polar coordinates and
$v_\phi=v_0+v$, where $v_0$ is the circular velocity at the Solar
radius, $r_0$. In this work we primarily consider a flat rotation
curve (RC; see, however \S \ref{sec:concl}), hence the derivatives of
$v_\phi$ in the above equations are identical to the derivatives of
$v$. $C$ describes the radial shear of the velocity field and $K$ its
divergence. For an axisymmetric Galaxy we expect vanishing values for
both $C$ and $K$\footnote{Note, however, that $C$ and $K$ would also
  be zero in the presence of non-axisymmetric structure if the Sun
  happened to be located on a symmetry axis.}. Whereas $C$ could be
derived from both radial velocities and proper motions, $K$ can only
be measured from radial velocities, in which case accurate distances
are also needed.

A problem with using proper motions data has been described by
OD03. The authors present an effect which arises from the longitudinal
variations of the mean stellar parallax caused by intrinsic density
inhomogeneities. Together with the reflex of the solar motion these
variations create contributions to the longitudinal proper motions
which are indistinguishable from the Oort constants at $\le$ 20\% of
their amplitude. OD03 corrected for the `mode-mixing' effect
described above, using the latitudinal proper motions.  The resulting
$C$ is found to vary approximately linearly with both color and
asymmetric drift (and thus mean age) from $C\approx0$ \ksk\ for blue
samples to $C\approx-10$ \ksk\ for late-type stars (see Figs. 6 and 9
in OD03).  Since $C$ is related to the radial streaming of stars, we
expect non-axisymmetric structure to mainly affect the low-dispersion
population which would result in the opposite behavior for $C$.  In
Paper I we showed that spiral structure failed to explain the observed
trend of $C$.

Note that the Oort constants are not constant unless they are measured
in the SN. Due to non-axisymmetries they may vary with the position in
the Galaxy.  Thus, the Oort constants have often been called the Oort
{\it functions}.

\section{The simulations}

We perform 2D test-particle simulations of an initially axisymmetric
exponential galactic disk. To reproduce the observed kinematics of the
Milky Way, we use disk parameters consistent with observations (see
Table \ref{table:par}).  A detailed description of our disk model and
simulation technique can be found in Paper I. We are interested in the
variation of $C$ with color ($B-V$), and asymmetric drift $v_a$. We
simulate variations with color by assuming that the velocity
dispersion increases from blue to red star samples.  For a cold disk
we start with an initial radial velocity dispersion
$\sigma_{u}=0.05v_0$ whereas for a hot disk we use
$\sigma_{u}=0.18v_0$.  The background axisymmetric potential due to
the disk and halo has the form $\Phi_0(r)=v_0^2\log(r)$, corresponding
to a flat RC.

\subsection{The bar potential}

We model the nonaxisymmetric potential perturbation due to the
Galactic bar as a pure quadrupole
\begin{eqnarray}
\Phi_{\rm b} = A_{\rm b}(\epsilon_{\rm b}) \cos[2(\phi-\Omega_{\rm b}
t)]\times        \\
\left\{
\begin{array}{cclcr}\nonumber
         \left(r_{\rm b}\over r\right)^3  &,&  r&\ge & r_{\rm b}    \\ 
       2-\left(r\over r_{\rm b}\right)^3  &,&  r&\le & r_{\rm b}
\end{array}
\right.
\end{eqnarray}
Here $A_{\rm b}(\epsilon_{\rm b})$ is the bar's gravitational
potential amplitude, identical to the same name parameter used by
\cite{dehnen00}; the strength is specified by $\epsilon_{\rm
  b}=-\alpha$ from the same paper.  The bar length is $r_{\rm
  b}=0.8r_{\rm cr}$ with $r_{\rm cr}$ the bar corotation radius. The
pattern speed, $\Omega_{\rm b}$ is kept constant.  We grow the bar by
linearly varying its amplitude, $\epsilon_{\rm b}$, from zero to its
maximum value in four bar rotation periods.  We present our results by
changing $\Omega_{\rm b}$ and keeping $r_0$ fixed.  The 2:1 outer
Lindblad resonance (OLR) with the bar is achieved when $\Omega_{\rm
  b}/\Omega_0=1+\kappa/2\approx1.7$, where $\kappa$ is the epicyclic
frequency. We examine a region of parameter space for a range of
pattern speeds placing the SN just outside the OLR. For a given
pattern speed one could obtain the ratio $r_0/r_{\rm OLR}$ through
$r_0/r_{\rm OLR}=\Omega_{\rm b}/\Omega_{\rm OLR}\approx\Omega_{\rm
  b}/1.7$.

\begin{figure}[]
\resizebox{\hsize}{!}{\includegraphics[clip=true]{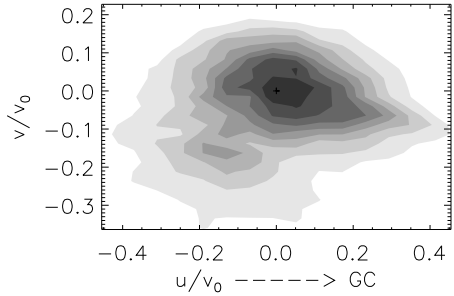}}
\caption{ \footnotesize 
  Simulated $u-v$ distribution for a bar pattern speed $\Omega_{\rm
    b}=1.87\Omega_0$, bar angle $\phi_0=35^\circ$ and a sample depth
  $d=r_0/40$. The initial velocity dispersion is $\sigma_{u}=0.18v_0$
  and the bar strength is $\epsilon_{\rm b}=-0.012$.  Contour levels
  are equally spaced. The clump identified with the Hercules stream is
  clearly discernible in the lower left portion of the plot. This
  figure is in agreement with the observed $f_0(u,v)$
  \citep{dehnen98,fux01} and simulated distribution functions of
  \cite{dehnen00} and \cite{fux01}.  }
\label{fig:uv}
\end{figure}

In contrast to \citet{dehnen00} and similar to \citet{fux01} and
\citet{muhlbauer03}, we integrate forward in time.  To allow for phase
mixing to complete after the bar reaches its maximum strength, we wait
for 10 bar rotations before we start recording the position and
velocity vectors. In order to improve statistics, positions and
velocities are time averaged for 10 bar periods.  After utilizing the
two-fold symmetry of our galaxy we end up with $\sim5\times10^5$
particles in a given simulated solar neighborhood with maximum radius
$d_{\rm{max}}=r_0/10$.

\section{Results}

First we show that we can reproduce the results of Dehnen and Fux for
the Hercules stream. We present a simulated $u-v$ velocity
distribution for $\Omega_{\rm b}=1.87\Omega_0$, $\phi_0=35^\circ$, and
a sample depth $\overline{d}=r_0/40$ in Fig. \ref{fig:uv}.  The
initial velocity dispersion is $\sigma_{u}=0.18v_0$.  This plot is
indeed in very good agreement with previous test-particle
\citep{dehnen00,fux01} and N-body \citep{fux01} simulations.

By a quantitative comparison of the observed with the simulated
distributions, \cite{dehnen00} deduced the Milky Way bar pattern speed
to be $\Omega_{\rm b}/\Omega_0=1.85\pm0.15$. In this calculation only
the local velocity field, i.e., the observed $u-v$ distribution, was
taken into account.  Now, if one could also relate the dynamical
effect of the bar to the derivatives of the velocities, an additional
constraint on bar parameters would be provided.  Since the gradients
of $u$ and $v$ are hard to measure directly, one needs an indirect way
of achieving this task. The obvious candidates, as anticipated, are
the Oort constants $C$ and $K$. Oort's $K$, however, is hard to
measure as mentioned above; hence only $C$, as estimated by OD03, will
be employed here.  By Fourier expansion of Eqs. \ref{eq:vel} (see
Paper I for details) we estimate $C$ from our numerically simulated
velocity distributions for the initially cold and hot disks.

\subsection{Variation of $C$ with bar pattern speed and orientation}
\label{sec:c}

\begin{figure*}[t!]
\resizebox{\hsize}{!}{\includegraphics[clip=true]{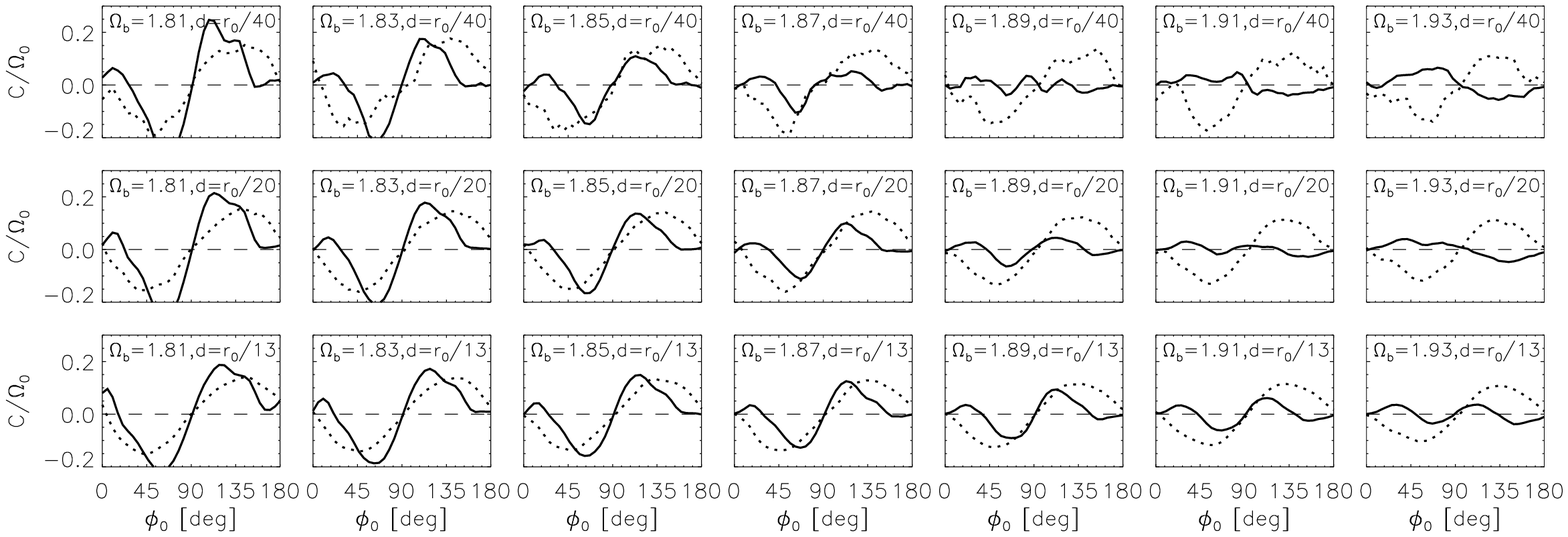}}
\caption{\footnotesize
  Each panel shows the variation of the Oort constant $C$ with bar
  angle $\phi_0$, for a simulation with the parameters given in Table
  \ref{table:par} and a particular bar pattern speed, $\Omega_{\rm
    b}$, and a mean sample depth, $\overline{d}$. Solid and dotted
  lines correspond to cold- and hot-disk values, respectively. Columns
  from left to right, show an increasing $\Omega_{\rm b}$ in units of
  $\Omega_0$. Note that the OLR is at $\Omega_{\rm
    OLR}\approx1.7$. Different rows present results from samples with
  different mean heliocentric distance, $\overline{d}$. Good matches
  to the observed trend in $C$ (vanishing value for the cold disk and
  a large negative for the hot one) are achieved for
  $20^\circ\leqslant\phi_0\leqslant45^\circ$ and
  $1.83\leqslant\Omega_{\rm b}/\Omega_0\leqslant1.91$.  }
\label{fig:c}
\end{figure*}

In Fig. \ref{fig:c} we present our results for $C$ as a function of
the bar angle, $\phi_0$ (the angle by which the Sun's azimuth lags the
bar's major axis). Each column shows a simulation with a different
pattern speed, indicated in each plot. Rows from top to bottom show
$C$ as calculated from samples at average heliocentric distances
corresponding to $\overline{d}=200, \overline{d}=400$, and
$\overline{d}=600$ pc, for a Solar radius $r_0=7.8$ kpc. Solid and
dotted lines represent the results for cold and hot disks,
respectively. The dashed lines indicates $C=0$. $C$ is presented in
units of $\Omega_0=v_0/r_0$. To make the discussion less cumbersome,
we write $C_{\rm h}$ and $C_{\rm c}$ to refer to the values for $C$ as
estimated from the hot and cold disks, respectively.

$C_{\rm h}$ (dotted lines in Fig. \ref{fig:c}) varies with galactic
azimuth as $C_{\rm h}(\phi_0)\sim\sin(2\phi_0)$ for all of the
$\Omega_{\rm b}$ values considered. On the other hand, the cold disk
values (solid line) exhibit different variations, depending on
$\Omega_{\rm b}$ or equivalently, on the ratio $r_0/r_{\rm OLR}$. Closer
to the OLR (left columns of Fig.  \ref{fig:c}), $C_{\rm c}(\phi_0)$
approaches the functional behavior of $C_{\rm h}(\phi_0)$. Away from
the OLR $C_{\rm c}(\phi_0)$ is shifted by $90^\circ$ compared to
$C_{\rm h}(\phi_0)$.  While both cold and hot disks yield in increase
in the amplitude of $C(\phi_0)$ as the pattern speed nears the OLR,
the effect on the $C_{\rm c}(\phi_0)$ is much stronger.  This is
consistent with our expectation that the cold disk is affected more by
the bar, especially near the OLR. While close to the OLR $|C_{\rm
  h}(\phi_0)|<|C_{\rm c}(\phi_0)|$, we observe the opposite behavior
away from it. This could be explained by the results of
\cite{muhlbauer03}, where they find that high velocity dispersion
stars tend to shift the `effective resonance' radially outwards.

According to OD03's `mode-mixing' corrected value, $C\approx0$ for
the cold sample and decreases to about $-10$ \ks\ for the hot
population. Hence we look for locations in Fig. \ref{fig:c} complying
with this requirement.  In the transition region,
$1.83\leqslant\Omega_{\rm b}/\Omega_0\leqslant1.91$, where the
function $C_{\rm c}(\phi_0)$ transforms into its negative, there are
specific angles which provide good matches for the observations, i.e.,
$C_{\rm c}(\phi_0)\approx0$ while $C_{\rm h}(\phi_0)$ is significantly
negative. We performed simulations with different pattern speeds in
the range $1.7\leqslant\Omega_b/\Omega_0\leqslant2.5$. It was found
that only in the transition region in Fig. \ref{fig:c} could one
achieve a satisfactory match to the observations. Although the left-
and right-most columns are nearly consistent with our requirement on
$C$, we reject these values of $\Omega_{\rm b}$ for the following
reason. OD03 estimated a large negative $C$ from both the reddest
main-sequence stars, which are nearby, and the more distant red
giants. This is inconsistent with the largest mean sample depth shown
in Fig. \ref{fig:c} with $\Omega_{\rm b}=1.81,1.93$.  We conclude that
the bar pattern speed must lie in the range
\begin{equation}
\Omega_{\rm b}/\Omega_0=1.87\pm0.04.
\end{equation}

The bar angle with respect to the Sun has been proposed, as derived
from IR photometry, to lie in the range $15^\circ-45^\circ$ with the
Sun lagging the bar major axis, whereas \cite{dehnen00} found that the
bar reproduced the $u$-anomaly for
$10^\circ\leqslant\phi_0\leqslant70^\circ$.  Examining
Fig. \ref{fig:c}, we find that our requirements on $C$ result in
constraining the bar angle in the range
$20^\circ\leqslant\phi_0\leqslant45^\circ$, depending on $\Omega_{\rm
  b}$.  This result for the dependence of pattern speed on bar angle
is in a very good agreement with Fig. 10 by \cite{dehnen00} which
shows a linear increase of the derived $\Omega_{\rm b}/\Omega_0$ as a
function of $\phi_0$. This is remarkable since \cite{dehnen00}
obtained his results in a completely different way than our method
here.

The Oort constants $A$ and $B$ were found to also be affected by the
bar, although to a much smaller extent. After correcting for
asymmetric drift, we found that for the cold disk the bar causes
$\Delta A=3.5\pm0.7$ and $\Delta B=3.0\pm0.6$ \ksk, while the hot disk
yielded $\Delta A=0.5\pm1.5$ and $\Delta B=0\pm1$ \ksk.  These values
are calculated for $r_0=7.8$ kpc and $v_0=220$ \ks.  Despite the bar's
effect on $C$, the hot population does provide better measurements for
$A$ and $B$.

\section{Conclusion}
\label{sec:concl}

We have shown that the Galactic bar can account for a trend seen in
measurements of Oort's $C$, namely a more negative $C$-value with
increasing velocity dispersion. At the same time we have improved the
measurement of the bar pattern speed, finding $\Omega_{\rm
  b}/\Omega_0=1.87\pm0.04$. In addition, the bar angle is found to be
in the range $20^\circ\leqslant\phi_0\leqslant45^\circ$. Our result
for $\Omega_{\rm b}$ lies well within the estimate by \cite{dehnen00}
and that by \cite{debattista02}, based on OH/IR star kinematics.  This
study provides an improvement on the measurement of the bar pattern
speed by a factor of $\sim4$ compared to previous work
\citep{dehnen00}.  The improved constraints on bar parameters should
be tested by, and will benefit, future studies of the bar structure in
the Galactic center region, that will become possible with future
radial velocity and proper motion studies (e.g., BRAVA,
\citealt{rich07}).

In addition to a flat RC we have also considered a power law initial
tangential velocity $v_\phi=v_0(r/r_0)^\beta$ with $\beta=0.1,-0.1$
corresponding to a rising and a declining RC, respectively. We found
that for $\beta=0.1$ no additional error in $\Omega_b$ is
introduced. However, for $\beta=-0.1$ we estimated $\Omega_{\rm
  b}/\Omega_0=1.85\pm0.06$.  From Fig. \ref{fig:c} we see that the
main source of error in the bar pattern speed is due to the
uncertainty in the bar angle. A recent work by \cite{rattenbury07}
used OGLE-II microlensing observations of red clump giants in the
Galactic bulge, to estimate a bar angle of $24^\circ-27^\circ$.  Using
this as an additional constraint we obtain $\Omega_{\rm
  b}/\Omega_0=1.84\pm0.01$.

While we account for the trend of increasingly negative $C$ with
increasing velocity dispersion, we fail to reproduce the size of the
$C$-value measured by OD03. The most negative value we predict is
about $-6$ while they measure $-10$ \ksk. We discuss possible reasons for
this discrepancy: (i) we expect that the magnitude of $C$ is related
to the fraction of SN stars composing the Hercules stream. It is
possible that during its formation the bar changed its pattern speed
forcing more stars to be trapped in the 2:1 OLR, thus increasing this
fraction. This would give rise to a more negative $C$ from hot stars,
while leaving the cold population unaffected; (ii) the numerical model
here uses a distance limited sample. Modeling a stellar population
with a magnitude limited sample is more appropriate to compare to the
observed measurements.  (iii) OD03's `mode-mixing' correction is only
valid if the Sun's motion in the $z$-direction were assumed
constant. This is invalidated if, for example, a local Galactic warp
were present.

\begin{acknowledgements}
We would like to thank Walter Dehnen for helpful comments. 
Support for this work was in part provided by NSF grant ASST-0406823, and
NASA grant No.~NNG04GM12G issued through the Origins of Solar Systems Program.
\end{acknowledgements}


\bibliographystyle{aa}

\end{document}